\documentclass[twocolumn,showpacs,preprintnumbers,amsmath,amssymb]{revtex4}

\usepackage{graphicx}
\begin{document}
\newcommand{\be}{\begin{equation}}
\newcommand{\ee}{\end{equation}}
\newcommand{\bea}{\begin{eqnarray}}
\newcommand{\eea}{\end{eqnarray}}
\newcommand{\bc}{\begin{center}}
\newcommand{\ec}{\end{center}}

\preprint{FT-07-03}

\title{The role of resonances  in $\rho^0 \rightarrow \pi^+ \pi^- \gamma$ }

\author{G. Toledo S\'anchez}

\affiliation{ Instituto de F\'{\i}sica, Universidad Nacional Aut\'onoma de M\'exico. M\'exico D. F. C.P. 04510}

\author{J. L. Garc\'{\i}a-Luna and V. Gonz\'alez-Enciso }

\affiliation{ Departamento de F\'{\i}sica, Centro Universitario de Ciencias Exactas e Ingenier\'{\i}as, Universidad de Guadalajara, Blvd. Marcelino Garc\'{\i}a Barrag\'an 1508, CP 44840, Guadalajara Jal., M\'exico}
\date{\today}
            
\begin{abstract}
We study the effect of the $\sigma(600)$ and $a_1(1260)$ resonances in the $\rho^0 \rightarrow \pi^+ \pi^- \gamma$ decay, within the meson dominance model.  Major effects are driven by the  mass and width parameters of the  $\sigma(600)$, and  the usually neglected contribution of the $a_1(1260)$, although small by itself, may become sizable through its interference with pion bremsstrahlung, and the proper relative sign can favor the central value of the experimental branching ratio.  We present a procedure, using the gauge invariant structure of the resonant amplitudes, to kinematically enhance the resonant effects in the angular and energy distribution of the photon. We also elaborate on the coupling constants involved.
\end{abstract}

\maketitle

\section{Introduction}
The $\rho^0$ meson can be produced via $e^+e^-$ annihilation. Dedicated experiments like KLOE, SND  and CMD-II  have performed extensive studies on the decay modes of this meson \cite{BRexperiment,experiments,experiments2, PDG06}. The measured  branching ratio of the $\rho^{0}\rightarrow \pi^{+} \pi^{-}\gamma$  radiative decay is $ 9.9\pm 1.6 \times 10^{-3}$ \cite{BRexperiment}. The theoretical approach is able to explain the experimental result, at the current precision, considering the radiation off pions alone. This can be estimated in a model independent way, predicting a branching ratio around one standard deviation above the central value \cite{singer,theoreticals, Bramon}. The maximum energy that the photon can carry out in this process is of $\omega=334.8 \ MeV$, which is large enough  to suggest that contributions beyond the soft photon approximation may be relevant in future accurate  measurements.\\
\indent The introduction of the subleading contributions is model dependent. If the scalar contribution is identified with the $\sigma(600)$ resonance  then both mass and width may be obtained by making a fit to the data. Still, the range of values obtained using this procedure are not as precise as the determination from other methods \cite{caprini, Aitala}. The other allowed contribution for the decay is of the axial-vector form. The coupling of the relevant amplitude is estimated in the chiral perturbation theory to be proportional to the parameters $L^r_9+L^r_{10} \simeq 1.4\times 10^{-3}$ \cite {chiral} and therefore is commonly disregarded in the analysis \cite{Bramon}. On the other hand, meson dominance identifies this contribution with the $a_1(1260)$ state which, at first, is far from being on-shell and then is neither taken into account.\\
\indent The improvement on the experimental precision will make it possible to distinguish the resonant contributions, and therefore a clear theoretical estimate will be important. In particular, we can ask whether such contributions can help to bring  the theoretical predictions closer to the central experimental value and if there is a window where the resonant contributions are more relevant. The quark structure and proper characterization of such states is also a matter of ongoing research \cite{structure,structureA,f980}.\\
\indent In this article we use the meson dominance model to provide a full account of these resonances. We perform a self-consistent determination of the effective coupling constants involved by reproducing observables in other decays, without invoking to one-loop corrections.  Since the radiation is dominated by pion bremsstrahlung, the resonant contribution upon its interference with it can become relevant. Here we compute  the  contributions to the branching ratio from the different sources which may also provide a hint to the relative sign of the $a_1(1260)$ amplitude by favoring the experimental central value. The di-pion invariant mass spectrum and  the angular and energy distribution of the photon are used to look for sensitivity to the resonances. We exploit kinematical configurations where for the latter  may lead to enhancements of the resonant contributions. The  $\sigma(600)$ mass and width effects are explored throughout the analysis.

       \begin{figure}
\includegraphics[width=2.5in,height=2.3in,angle=0]{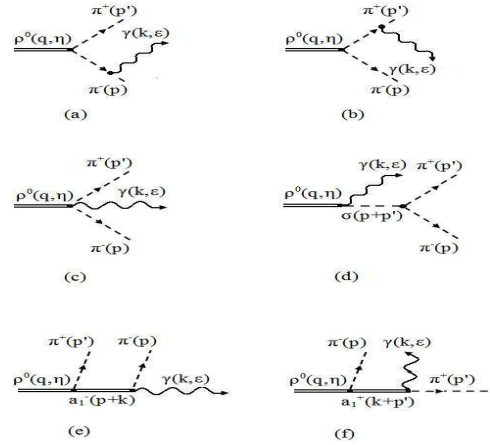}
   \caption{ Feynman diagrams that contribute to the decay  $
\rho^{0}\rightarrow \pi^{+} \pi^{-} \gamma$.}
   \label{dfeynman2}
   \end{figure}

\section{ Decay amplitudes}
Let us state the conventions for the process $\rho^0(q,\eta)\rightarrow  \ \pi^+(p^{'}) \pi^-(p) \     \gamma(k,\epsilon)$; $q$,  $p^{'}$, $p$ and  $k$ are the corresponding 4-momenta and, $\eta$ and $\epsilon$ are the polarization tensors of the vector meson and the photon respectively.
The Feynman diagrams contributing to the decay are shown in Figure  (\ref{dfeynman2}). Diagrams \ref{dfeynman2}a, \ref{dfeynman2}b and \ref{dfeynman2}c are model independent  and correspond to the Low amplitude \cite{Low}. Diagrams \ref{dfeynman2}d, \ref{dfeynman2}e and \ref{dfeynman2}f are model dependent and, in the meson dominance model, include the intermediate $\sigma$ and $a_1$ resonances. We will use the dependence on these states to label the corresponding amplitudes. 
The intermediate state $f(980)$ is not taken into account since  it is considered to be dominated by the $KK$ channel \cite{f980}. Therefore, the total amplitude can be written as:
\be
\label{ec:eje4.6a}
          \mathcal{M}_{T}=
          \mathcal{M}_{L}+ \mathcal{M}_\sigma+
\mathcal{M}_{a_1^{+}} +  \mathcal{M}_{a_1^-},
\ee
where the Low amplitude \cite{singer} $ \mathcal{M}_{L}=\mathcal{M}_{a}+\mathcal{M}_{b}+\mathcal{M}_{c},$ can be split into two parts, owing to its dependence on the photon energy:  $\mathcal{M}_{L}=\mathcal{M}_{e}(\omega^{-1}) + \mathcal{M}_{0}(\omega^{0})$, where 
 \be
 \mathcal{M}_{e}=2ie \ g_{\rho \pi\pi} \ p^{'}
\cdot \eta\
   L \cdot \epsilon^*,
\ee
\be
\mathcal{M}_{0} =2ie \ g_{\rho \pi\pi}
\left[\epsilon^* \cdot \eta- \frac{p^{'} \cdot \epsilon^*\ k\cdot \eta}
{p^{'} \cdot k}
   \right]
\ee
and $L^{\nu}=p^\nu/p\cdot k -p'^\nu/p' \cdot k$ satisfies $L \cdot k = 0$. Therefore $M_L$ is explicitly gauge invariant.   $g_{\rho \pi\pi}$ is the coupling constant between the rho and the pions, as indicated by the subindex. The amplitudes for figures \ref{dfeynman2}d,  \ref{dfeynman2}e and \ref{dfeynman2}f are  gauge invariant by themselves  and are given by:
\be
\label{ec:eje4.5a}
\mathcal{M}_\sigma= -i e g_{\rho \sigma \gamma} \ g_{\sigma \pi \pi}
    \left[ \frac{q \cdot k \ \epsilon^* \cdot \eta - q \cdot \epsilon^* \
k \cdot \eta}
    {(p + p^{'})^{2}- m_{\sigma}^{2}+ i m_{\sigma}
    \Gamma_{\sigma}}\right],
\ee

      \begin{eqnarray}\label{ec:eje4.5b}
\mathcal{M}_{a_1^{+}}=\frac{i \ g_{a_1\pi\gamma} \ g_{a_{1}\rho\pi} p^{'} \cdot k (k+ p^{'})\cdot q}
{(k + p^{'})^{2} - m_{a_1}^{2} + i m_{a_{1}} \Gamma_{a_{1}}}\epsilon^{*\lambda} \eta^{\theta} \times  \nonumber \\ 
 \left(-g^{\nu\lambda} +
     \frac{p^{'\lambda} k^{\nu}}{p^{'} \cdot k}\right)
      \left(g^{\theta\nu}- \frac{(k + p^{'})^\theta q^{\nu}}{(k+ p^{'})\cdot
q}\right) ,
\end{eqnarray}

      \begin{eqnarray}\label{ec:eje4.5c}
\mathcal{M}_{a_1^{-}}=\frac{i \ g_{a_1\pi\gamma} \ g_{a_{1}\rho\pi} p \cdot k (k+ p)\cdot q}
{(k + p)^{2} - m_{a_1}^{2} + i m_{a_{1}} \Gamma_{a_{1}}}\epsilon^{*\lambda} \eta^{\theta} \times  \nonumber \\ 
 \left(-g^{\nu\lambda} +
     \frac{p^{\lambda} k^{\nu}}{p \cdot k}\right)
      \left(g^{\theta\nu}- \frac{(k + p)^\theta q^{\nu}}{(k+ p)\cdot
q}\right),
\end{eqnarray}
respectively, where  $g_{\sigma \pi \pi}$, $g_{a_{1} \pi \gamma}$ and $ g_{a_{1} \rho \pi}$ are effective coupling constants. $M_\sigma$ ($\Gamma_{\sigma}$) and  $m_{a_1}$ ($\Gamma_{a_1}$) are the corresponding particle masses (full widths). The resonant propagators have been assumed to be of the complex mass form \cite{complexmass}, which corresponds to replacing $m^2 \rightarrow m^2-im\Gamma$ in the non-resonant propagator.

\section{Squared amplitudes and interferences}
We choose  $p \cdot k$ and $p^{'} \cdot k$ as the independent variables to make the dependence on the photon energy explicit. We list below some of the relevant squared amplitudes and interferences:
      \begin{equation}\label{ec:eje4.a}
     |\mathcal{M}_{e}|^2=16 \pi  \alpha   g_{\rho \pi \pi }^2 L^2
\left( m_{\pi}^2 - \frac{M_{\rho}^2}{4}-\frac{
     (p \cdot k)^2}{M_{\rho}^2}\right),
\end{equation}

      \begin{eqnarray}\label{ec:eje4.b}
     |\mathcal{M}_{0}|^2 &=& 16\pi \alpha  g_{\rho \pi \pi}^2 
\left(1+
\frac{(p \cdot k+ p^{'} \cdot k)}{ M_{\rho}^2 p^{'} \cdot k} \times
\right. \nonumber  \\ 
 && \left. \left( M_\rho^2 - m_{\pi}^2 -
     \frac{p \cdot k \ m_{\pi}^2}{p^{'} \cdot k} - 
     2 p \cdot k \right) \right) ,
\end{eqnarray}

\begin{eqnarray}\label{ec:eje4.c}
  2Re \ \mathcal{M}_{e} \mathcal{M}_{\sigma}^\dagger=16 \pi\alpha \ g_{\rho \pi
\pi } \ g_{\rho\sigma\gamma}\ g_{\sigma\pi\pi}p \cdot k p^{'} \cdot k L^2 
    \times  \\
  \left(\frac{M_{\rho}^2-
    m_{\sigma}^2-2 q\cdot k}
{[M_{\rho}^2-m_{\sigma}^2-2q\cdot k]^2
    + m_{\sigma}^2 \Gamma_{\sigma}^2}\right),\nonumber
\end{eqnarray}

\begin{eqnarray}\label{ec:eje4.d}
   2Re \ \mathcal{M}_{e} \mathcal{M}_{a_1^+}^\dagger=4 e \ g_{\rho \pi
\pi} \ g_{a_{1}\pi \gamma}\ g_{a_{1}\rho\pi}p \cdot k p^{'} \cdot k L^2
 \times  \\
   \left(\frac{[m_{\pi}^2+p^{'}
\cdot k ] [m_{\pi}^2- m_{a1}^2 + 2 p^{'} \cdot k]}
{[m_{\pi}^2 +
2p^{'} \cdot k- m_{a1}^2]^2 + m_{a1}^2
     \Gamma_{a1}^{2}} \right), \nonumber
\end{eqnarray}

      \begin{eqnarray}\label{ec:eje4.e}
     2Re \ \mathcal{M}_{e} \mathcal{M}_{a_1^-}^\dagger=4 e  \ g_{\rho \pi\pi} \ g_{a_{1}\pi \gamma}\  g_{a_{1}\rho\pi}p \cdot k p^{'} \cdot k L^{2}
     \times  \\
    \left( \frac{[m_{\pi}^{2}+2 p \cdot
k][m_{\pi}^{2}- m_{a1}^{2} + 2 p \cdot k]}
 {[m_{\pi}^{2} + 2p \cdot k-
m_{a1}^{2}]^{2} + m_{a1}^{2}
     \Gamma_{a1}^{2}} \right). \nonumber
\end{eqnarray}

We have not written the square of the model dependent amplitudes and the remaining interferences but they are actually taken into account in the calculation. The Low interferences of order  $\omega^{-1}$ are null in accordance with the Burnett-Kroll theorem \cite{Burnett} and a term proportional to $L^2$ from the $ \mathcal{M}_{e}$  $ \mathcal{M}_{0}^\dagger$ interference was absorbed into eqn. (\ref{ec:eje4.a}).  By inspection of eqns. (\ref{ec:eje4.a} - \ref{ec:eje4.e}) we observe that, in addition to $|\mathcal{M}_{e}|^{2}$, all the interferences are proportional to $L^{2}$, a property we showed in a previous work to hold whenever there is an interference between the electric charge radiation and any gauge invariant amplitude \cite{Castro}. This property will allow us to kinematically enhance the model dependent contribution by properly choosing the region where $L^2$ is maximum \cite{toledoPRD02}. Although this enhancement is also promoted to the dominant electric charge contribution, the latter  receives a natural suppression as the photon energy increases, owing to the $\omega^{-2}$ dependence,   while the model dependent ones are of order $\omega^0$ and higher.

\begin{table}
\begin{center}
\begin{tabular}{|l|l|l|l|l|l|}\hline

 $M_\sigma$ & $\Gamma_\sigma$ &  Sigma    & $Low + \sigma$  & Tot. (+)& Tot. (-)\\

MeV     &  MeV      & $10^{-5}$     & $10^{-3}$      & $10^{-3}$     & $10^{-3}$  \\

\hline

  500  &   500  &    1.7    &  11.56  &   $11.64\pm 0.03$  &    $11.49\pm 0.01$\\
  500  &   450  &    2.0    &  11.57  &     11.64  &    11.50\\
  500  &   400  &    2.4    &  11.57  &     11.64  &    11.50\\
  500  &   350  &    2.9    &  11.58  &     11.65  &    11.50\\
  450  &   500  &   -5.8    &  11.49  &     11.56  &    11.42\\
  450  &   450  &   -6.1    &  11.49  &     11.56  &    11.42\\
  450  &   400  &   -6.4    &  11.48  &     11.55  &    11.41\\
  450  &   350  &   -6.5    &  11.48  &     11.50  &    11.41\\
  400  &   500  &   -11.7   &  11.43  &     11.49  &    11.36\\
  400  &   450  &   -12.4   &  11.42  &     11.49  &    11.35\\
  400  &   400  &   -13.0   &  11.42  &     11.49  &    11.35\\
  400  &   350  &   -13.7   &  11.41  &     11.48  &    11.34\\
  350  &   500  &   -17.0   &  11.38  &     11.45  &    11.31\\
  350  &   450  &   -17.9   &  11.37  &     11.44  &    11.30\\
  350  &   400  &   -18.8   &  11.36  &     11.43  &    11.29\\
  350  &   350  &   -19.7   &  11.35  &     $11.42\pm 0.05$  &   $11.28 \pm 0.07$\\

\hline
  \end{tabular}
   \end{center}\caption{Branching ratios from several contributions for a set of values of the $\sigma$ parameters and a cut off on the photon energy of 50 $MeV$. $Low=11.547 \times 10^{-3}$. To have an idea of the effect from the uncertainties in the coupling constants and $a_1$ mass and width, we have included error bars in the first and last row. See text for details.}
   \label{BR}
\end{table}

\section{Results}
In order to make an estimate of the observables, we  first address the problem of finding the proper values of the coupling constants.
The  $g_{\rho\pi\pi}$  and $g_{\sigma\pi\pi}$  couplings can be written in terms of masses and widths as:
\be
  g_{\rho\pi\pi}^2=\frac{48  \pi 
\Gamma_\rho}{M_\rho} \left(1-\frac{4
m^{2}_{\pi}}{M^{2}_{\rho}}\right)^{-3/2}=(6.01)^2.
\ee
\be
 g_{\sigma\pi\pi}^2=\frac{32 \pi  \Gamma_{\sigma}
 m_{\sigma}}{3\sqrt{1-\frac{4
m^{2}_{\pi}}{m^{2}_{\sigma}}}}
\ee
where we have used for definiteness  $\Gamma_{\rho}=150.7 MeV$ and $ m_{\rho}=775 MeV$.\\
For the $g_{\rho\sigma\gamma}$ coupling, we use the expression depending on the radiative width  $\Gamma_{\rho\sigma\gamma}$ whose value was estimated  to be  $\Gamma_{\rho\sigma\gamma}=0.23\pm 0.47 \ keV$ or $ 17\pm 4 \ keV$ \cite{Black}.
\be
 g_{\rho\sigma\gamma}^2=\frac{3}{\alpha} \Gamma_{\rho\sigma\gamma}
\left(\frac{2 M_{\rho}}{M_{\rho}^{2}-m^{2}_{\sigma}}\right)^3.\label{grsg}
\ee
Taking the prediction for  $\Gamma_{\rho\sigma\gamma}=17\pm 4 \ keV $ and varying $ m_{\sigma}=350-500 MeV$ produces $  g_{\rho\sigma\gamma}^2=(1.82 \ to \ 7.46) \times 10^{-7} \ MeV^{-2}$. To compare  with the one obtained in \cite {gokalpPRD62} we re-parameterize the coupling by a $M_\rho$ factor ( $ g_{\rho\sigma\gamma}\rightarrow  \hat{g}_{\rho\sigma\gamma}=M_\rho g_{\rho\sigma\gamma}\rightarrow 0.33 \ to \ 0.67$). Therefore, it is one order of magnitude smaller than the extracted in \cite{gokalpPRD62} which lies in the region of $|6 \ to \ 7|$, and produces an overestimate of  the branching ratio for $\rho^0 \rightarrow \sigma\pi \rightarrow \pi^0 \pi^0 \gamma$ \cite{gokalpPRD67}. Our equation (\ref{grsg}) is valid in the limit of $\Gamma_{\sigma}=0$. In reference \cite{coupling} the coupling was computed including the width effect and relying on the experimental value for $\Gamma(\rho^0 \rightarrow \sigma\pi \rightarrow \pi^0 \pi^0 \gamma)=(2.9^{+1.4}_{-1.2}\pm0.6) \ keV$ \cite{experiments2}. Using theirs master equation  and varying  $m_\sigma$ and $\Gamma_\sigma$ from 350 to 500 $MeV$ produces  $ \hat{g}_{\rho\sigma\gamma}= 0.22 \ to \ 0.89$, which is of the same order than our value, although in a different approximation. Therefore the use of either values will produce similar results in the range we are exploring.\\
An estimate of the $g_{a_{1}\rho\pi}$ coupling can be obtained from the $a_1(q,\eta) \rightarrow \rho(k,\epsilon) \pi$ decay. The amplitude in its simplest on-shell form  becomes \cite{Isgur}:
\be
{\cal M}(a_1 \rightarrow \rho \pi)=  f_{a_{1}\rho\pi}(\eta \cdot \epsilon^* - k\cdot \eta \frac{q\cdot \epsilon^*}{k\cdot q})
\ee
 Our coupling is related to this by $ g_{a_{1}\rho\pi}=-f_{a_{1}\rho\pi}/ k\cdot q$. Assuming that the full width  is dominated by the $\rho \pi$ channel, the coupling is
\be
 f_{a_{1}\rho\pi}^2= \frac{12 \pi m_{a_1}^3 \Gamma_{a_1}}
{ [\rho^- \pi^0] \ + [ \rho^0 \pi^-]}=(3.7 \ to \ 5.9  \ GeV)^2
\ee
where $[\rho \pi]\equiv [1+M_{\rho}^2 m^2_{a_1}/(2(k\cdot q)^2)]\sqrt{(k\cdot q)^2-M_{\rho}^2 m^2_{a_1}}$ and $k\cdot q=(m^2_{a_1} + M_{\rho}^2- m_{\pi}^2)/2$, and we have used  $m^2_{a_1}=1230 \pm 40$ and $\Gamma_{a_1}=250-600 MeV$. This compares  well with  the prediction of $ f_{a_{1}\rho\pi}=4.8 \ GeV$ in the quark model \cite{Isgur}.\\
Finally, using  vector meson dominance arguments \cite{Gell-Mann62}, we can relate the $g_{a_{1}\rho\pi}$ and $ g_{a_{1}\pi \gamma}$ as follows:
$ g_{a_{1}\pi \gamma}=  e g_{a_{1}\rho\pi}/\gamma_\rho $, where $\gamma_\rho = 2 \alpha \sqrt{ \pi M_\rho / 3 \Gamma(\rho^0 \rightarrow e^+ e^-)}=5.012$.\\
Errors in the observables coming from the uncertainties on the coupling values will be specified below.\\

\begin{figure}[t]
\includegraphics[width=2.5in,height=2in,angle=0]{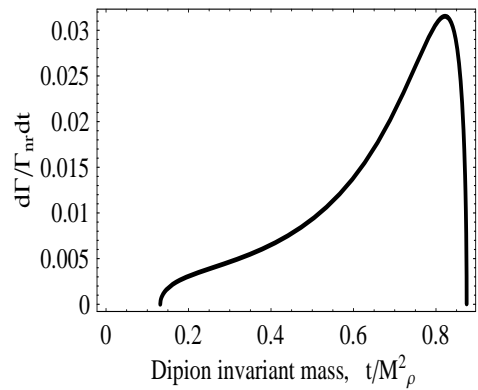}
\caption{Di-pion invariant mass due to the pions Bremsstrahlung for the $\rho^0 \rightarrow \pi^+ \pi^- \gamma$ decay.}
\label{dipion}
\end{figure}

The radiative decay width can be computed using the expressions for the squared amplitudes and interferences separately. In table \ref{BR}, we present the different contributions to the branching ratio for a set of values for the mass and width of the $\sigma(600)$. We have introduced a $50 \ MeV$ cut in the photon energy to avoid the infrared divergence, then its allowed region is:
$ \omega_{cut} \leq \omega \leq (M_{\rho}^{2}- 4m_{\pi}^2)/2M_{\rho}=334.8 \ MeV$. The upper value requires to go beyond the soft photon approximation, given by the Low amplitude. It is worth to mention that the higher the cut in the photon energy the better the sensitivity to the resonant parameters. Here we stick to this value to compare with the available data.  The Low amplitude contribution to the branching ratio is $Low = 11.547\times 10^{-3} $,  about one standard deviation above the experimental value of $ 9.9\pm 1.6 \times 10^{-3}$ \cite{BRexperiment}, for the same cut off. The column labeled ``Sigma'' corresponds to the contribution  from the $\sigma$ amplitude itself and its interference with the Low amplitude. This becomes larger for smaller values of the $\sigma$ parameters and can even flip the sign. Column labeled ``Low +sigma'' is the sum of the previous column plus the $Low$ contribution. Columns labeled ``Tot(+)'' and ``Tot(-)'' correspond to the total branching ratio when the $a_1$  interferences with the Low amplitude is also included. Depending on the sign of the $a_1$ amplitudes the contribution can be either $a_1(+)=6.8\times 10^{-5}$ or $a_1(-)=-6.6 \times 10^{-5}$.\\
\indent From table \ref{BR}, we observe that the inclusion of the resonances can help to bring the prediction  closer to the central experimental value. In particular the sign on the $a_1$ amplitude can either improve or worsen the agreement. Still, the major effect is driven only by the $m_\sigma $ and $ \Gamma_\sigma$ parameters. To have an idea of the effect from the uncertainties in the coupling constants discussed previously and $a_1$ mass and width, we have included error bars in the first and last row of Table \ref{BR}. The contributions from the $a_1$ itself is of order of $10^{-6}$ and neglected in the results.\\
\indent The di-pion invariant mass  is interesting on its own and the experiments usually report it as the main observable. However, in our case the individual effects from the interferences between the electric radiation and the resonances are very mild, as can be expected from the results for the branching ratios. Just for illustration, in Figure \ref{dipion} we plot the corresponding Low amplitude, which dominates this observable.

       \begin{figure}
\includegraphics[width=2.5in,height=2in,angle=0]{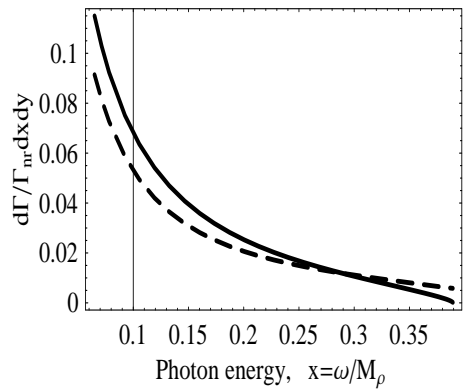}
\caption{ Photon energy spectrum due to the Low emission for $\theta=85^{\circ}$ and $ 5^{\circ}$ (solid and dashed lines respectively).}
   \label{electric}
   \end{figure}

\begin{figure}
\includegraphics[width=2.5in,height=2in,angle=0]{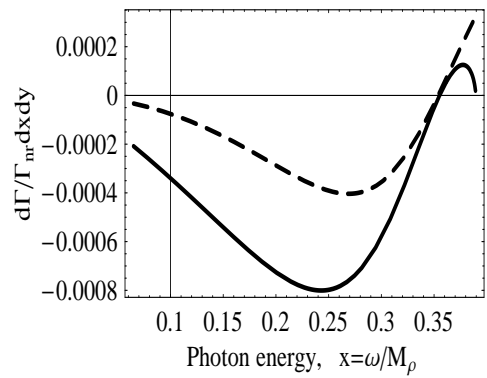}
\caption{ Photon energy spectrum due to the interference between electric and $\sigma$ emission for $\theta=85^{\circ}$ and $ 5^{\circ}$ (solid and dashed lines respectively). Here $m_\sigma=\Gamma_\sigma=400 \ MeV$. }
   \label{electricsigma}
   \end{figure}

\begin{figure}[t]
 \includegraphics[width=2.5in,height=2in,angle=0]{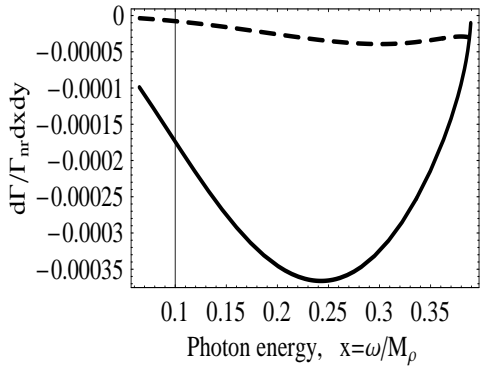}
 \caption{ Photon energy spectrum  due to the interference between electric and $a_1$ emission for $\theta=85^{\circ}$ and $ 5^{\circ}$ (solid and dashed lines respectively).}
   \label{electrica1}
   \end{figure}

The fact that all the leading interferences are proportional to 
$L^2$, allows to look for an enhancement of the effects from the resonances by choosing the kinematical configuration where $L^2$ is  maximum. In the $\rho^0$ restframe it can be written as  $L^2 \propto 1-cos^2 \theta$ where $\theta$ is the angle between the photon and $\pi^-$ 3-momenta. In addition, the dependence on the photon energy of the interferences is of order  $\omega^0$ while the dominant Low contribution is of order   $\omega^{-2}$. This suggests that an  appropriate observable could be the angular and energy distribution of the photon. 
\begin{eqnarray}\label{angular}
   d\Gamma = \sum_{E= E^+,E^-}\frac{|\overline{\mathcal{M}}|^{2}}{(2\pi)^{5}}
\frac{M_{\rho}}{8} \frac{\sqrt{E^2-m_\pi^2}}{\mid
F^{'}(E)\mid} x dx dy,
\end{eqnarray}
where, $E^\pm$ labels the roots of $ F(E)= M_{\rho}^{2} - 2M_{\rho}E - 2M_{\rho}\omega + 2(E w -  p \omega cos \theta )$, in the $\rho^0$ restframe, and we have introduced the dimensionless variables $y\equiv cos \theta$ and $x \equiv \omega/M_\rho$.  The maximum value  for $\omega$ is $(M_\rho^2 -2 m_\pi M_\rho)/2(M_\rho-m_\pi)$.
In Figures \ref{electric}, \ref{electricsigma} and \ref{electrica1} we plot $d\Gamma/\Gamma_{nr} dydx $ (normalized to the non-radiative width, $\Gamma_{nr}$) for two angles of the photon emission, $\theta=85^{\circ}$ and $ 5^{\circ}$ (solid and dashed lines respectively).  Figure \ref{electric} includes the Low contribution,  Figure \ref{electricsigma} the interference between Low and $\sigma$, and  Figure \ref{electrica1} the interference between Low and $a_1$ emission. Here we have used $M_\sigma=\Gamma_\sigma=400 \ MeV$.
We can observe that, although small, the contributions from the resonances can be strongly enhanced by choosing the proper angle off emission, about 50\%  for the $\sigma$ and upto 85\% for the $a_1$ in the current set up of relative angles. This enhancement is not promoted to the Low emission at the same  proportion,  which is mildly affected and even suppressed for large values of the photon energy as expected (Figure \ref{electric}). The total photon energy and angular spectrum is certainly dominated by the Low emission but this is free of relevant theoretical uncertainties and can be safely removed from data.

\section{Conclusions}
We have studied  the  $\rho^{0}\longrightarrow \pi^{+} \pi^{-} \gamma$ decay in a self-consistent approach based on the meson dominance model, and included both the $\sigma(600)$ and $a_1(1260)$ resonances.  We determined the corresponding coupling constants involved and, in particular, the value for   $  g_{\rho\sigma\gamma}$  is one order of magnitude smaller than the one extracted  in previous studies \cite{gokalpPRD62} but similar to the estimation made in \cite{coupling} for the particular range of parameters we did explore. Rigorously one must use the coupling from the latter where $\Gamma_\sigma$ is taken into account.\\
\indent We identified the different contributions to the branching ratio,  dominated by the pion bremsstrahlung whose contribution alone lies one standard deviation above the experimental central value (measured with a precision of about 16\%). The resonant contributions upon interference with the Low radiation, although small, can be of relevance for future accurate measurements of the branching ratio and be sensitive to the resonance parameters. In fact, this would provide a hint on the relative sign of the axial amplitudes by requiring the theoretical prediction to lie closer to the central experimental value. In order to distinguish the effects coming from the $\sigma(600)$ parameters a precision smaller than $5 \%$ is required, while to be sensitive to the $a_1$ parameters at least the one percent level is required.\\
\indent By another hand, the di-pion invariant mass spectrum  was computed and shown to be saturated by the pion Bremsstrahlung.\\
\indent Exploiting the structure of the leading interferences, we tuned a kinematical configuration where  resonant contributions can be enhanced, namely the photon angular and energy spectrum. In particular   the effects can be enhanced for quasi-transversal emission  compared to quasi-collinear emission of photons, with respect to  the $\pi^-$ 3-momentum. Our treatment is useful for looking for  enhancements of the resonances in decays of the form  $\rho^0 \rightarrow \pi^+ \pi^- \gamma$, since it exploits the radiation structure of the external charged particles and can serve as a complement to estimates from decays of the form  $\rho^0 \rightarrow \pi^0 \pi^0 \gamma$ which are mainly driven by model dependent contributions and where charged particles contributes only through loops and therefore our approach can not be applied.

{\bf Acknowledgments} We are  grateful to G. L\'opez Castro and Jens Erler for very useful observations. This work was partially supported by Conacyt M\'exico, under grant 42026-F.

\end{document}